\documentclass[preprint,prc,aps,epsfig]{revtex4}
\usepackage{psfig}

\def\beq{\begin{equation}}
\def\eeq{\end{equation}}
\def\beqa{\begin{eqnarray}}
\def\eeqa{\end{eqnarray}}
\def\MeV{\nobreak\,\mbox{MeV}}
\def\GeV{\nobreak\,\mbox{GeV}}

\def\mpi{m_\pi}
\def\md{m_D}
\def\mds{m_{D^*}}
\def\mpsi{m_\psi}

\newcommand{\fslash}[1]{\ooalign{\hfil/\hfil\crcr$#1$}}

\begin{document}

\title{Progress in the determination of the $J/\psi-\pi$ cross section}

\author{Francisco O. Dur\~aes$^1$, Hungchong Kim$^2$,
Su Houng Lee$^{2,3}$, Fernando S. Navarra$^1$ and Marina Nielsen$^1$}
\affiliation{$^1$Instituto de F\'{\i}sica, Universidade de S\~{a}o Paulo, 
C.P. 66318,  05315-970 S\~{a}o Paulo, SP, Brazil\\
$^2$Institute of Physics and Applied Physics, Yonsei University,
Seoul 120-749, Korea\\
$^3$Cyclotron Institute, Texas A\&M University, College Station, 
TX 77843-3366, USA.}

\begin{abstract}
Improving previous calculations, we compute the $J/\psi~\pi\rightarrow 
\mbox{charmed mesons}$ cross section using QCD sum rules. Our sum rules
for the $J/\psi~\pi\rightarrow \bar{D}~D^*$, $D~\bar{D}^*$, 
${\bar D}^*~D^*$ and ${\bar D}~D$ hadronic matrix elements are constructed by 
using vaccum-pion correlation functions, and we work up to twist-4
in the soft-pion limit.  
Our results suggest that,  using meson exchange models is perfectly
acceptable, provided that they include form factors and  that they  
respect chiral symmetry. 
After doing a thermal average we get $\langle\sigma^{\pi J/\psi} v\rangle\sim
0.3$ mb at $T=150\MeV$.
\end{abstract}

\maketitle

\vspace{1cm}
\section{Introduction}

For a long time charmonium suppression has been considered as one of the best 
signatures of 
quark gluon plasma (QGP) formation \cite{ma86}. Recently this belief was 
questioned by some 
works. 
Detailed simulations \cite{thews} 
of a population of $c - \overline{c}$ pairs traversing the plasma suggested 
that, given the 
large number of such pairs, the recombination effect of the pairs into  
charmonium Coulomb bound states is 
non-negligible and can even  lead to an enhancement of $J/\psi$ production. 
This conclusion 
received support from  the calculations of \cite{rapp}. Taking the existing 
calculations seriously,  
it is no longer clear that an overall suppression of the number of $J/\psi$'s 
will be a signature of QGP.  
A more complex pattern can emerge, with suppression in some regions of the 
phase space and enhancement in others \cite{huf,dnn}. Whatever the new QGP 
signature (involving charm) turns 
out to be, it is necessary to understand better the $J/\psi$ dissociation 
mechanism by 
collisions with comoving hadrons. 

Since there is no direct experimental information on $J/\psi$ absorption cross 
sections by hadrons, several theoretical approaches have been proposed to 
estimate their values.
In order to elaborate a theoretical description of the phenomenon, we have 
first to choose the relevant degrees of freedom. Already at this point no 
consensus has been found. Some 
approaches were based on charm quark-antiquark dipoles interacting with the 
gluons of a larger 
(hadron target) dipole \cite{bhp,kha2,lo} or quark exchange between two 
(hadronic) bags 
\cite{wongs,mbq},  whereas 
other works used the meson exchange mechanism 
\cite{mamu98,osl,haglin,linko,ikbb,nnr}. 
In this case it is not easy to decide 
in favor of quarks or hadrons because we are dealing with charm quark bound 
states, which are small and massive enough to make perturbation theory 
meaningful, but not small enough to make 
non-perturbative effects negligible. Charmonium is ``the borderguard of the 
mysterious border of perturbative world of quarks and gluons and the 
non-perturbative world of hadrons'' 
\cite{kardok}.

In principle, different approaches apply to different energy regimes and we 
might think that  
at lower energies  we can use quark-interchange
models \cite{wongs,mbq} or meson exchange models 
\cite{mamu98,osl,haglin,linko,ikbb,nnr} and, at higher energies we can use 
perturbative QCD   \cite{bhp,kha2,lo}.  However, even at low energies, the 
short distance aspects may become dominant and spoil  a non-perturbative 
description. 
In a similar way, non-perturbative effects may be important even at very high 
energies \cite{hdnnr}.

Inspite of the difficulties, 
some progress has  been achieved. This can be best realized if we  compare our 
knowledge 
on the subject today with what we knew a few years ago,  
described by B. Mueller (in 1999)  
\cite{muel} as ``...the state of the theory of interactions 
between $J/\psi$ and light hadrons is embarrassing. Only three serious 
calculations exist (after more than 
10 years of intense discussion about this issue!) and their results differ by 
at least two orders of magnitude 
in the relevant energy range. There is a lot to do for those who would like  
to make a serious contribution to 
an important topic''. In the subsequent three  years about 30 papers on this 
subject appeared and now 
the situation is much better, at least in what concerns the determination of 
the order of magnitude, which, 
as it will be discussed below, in the case of the $J/\psi$ pion interaction, 
is determined to be 
$ 1 \, < \,  \sigma_{J/\psi -  \pi} \, < \, 10 $ mb in the energy region close 
to the open charm production threshold.

One of the main things that we have learned is the near-threshold behavior of 
$\sigma_{J/\psi-\pi}$. This is quite relevant because in a hadron gas at 
temperatures of $100 - 300$ MeV most of the $J/\psi-\pi$ 
interactions occur at relatively low energies, barely sufficient to dissociate 
the charmonium 
state. In some calculations a rapid growth of the cross sections with the 
energy was 
found \cite{haglin,linko,osl}. 
This behavior was criticized and considered to be incompatible with empirical 
information 
extracted from  $J/\psi$ photoproduction \cite{zinovjev}.  This criticism, 
however, made use of the vector 
meson dominance hypothesis (VDM), which, in the case of charm, is rather 
questionable \cite{hufkop}. The 
introduction of form factors in the effective Lagrangian approach, while 
reducing the order 
of magnitude of the cross section, did not change this fast growing trend 
around the threshold.  Later, again in 
the context of meson exchange models, it was established \cite{nnr} that the 
correct implementation of
chiral symmetry prevents the cross section from rising steeply around the 
threshold. In a different approach, with QCD sum rules (QCDSR) \cite{nnmk}, 
the behavior found in \cite{nnr} was confirmed and in the present work, with 
improved QCDSR, we confirm again the smooth threshold 
behavior. We, thus, believe that this question has been answered.

Another, phenomenologically less important, but conceptually interesting issue 
is the energy 
dependence  in the region far from threshold. Results obtained with the 
non-relativistic quark model \cite{wongs} indicated a rapidly falling cross 
section. This behavior is due to the gaussian tail of the quark wave 
functions used in the quark exchange model.   
This result of the quark model approach could be mimicked within chiral meson 
Lagrangian 
approaches with the introduction of  $\sqrt{s}$ dependent form
factors \cite{oslw,ikbb}. For $J/\psi-N$ interactions, 
it was found in ref. \cite{lo} that this behaviour depends ultimately on the  
gluon distribution 
in the proton at low $x$. In the case of $J/\psi-N$, for certain 
parametrizations of the gluon
density one could find a falling trend for the cross section \cite{lo}, 
but no definite 
conclusion could 
yet be drawn.  

If the $J/\psi$ is treated as an ordinary hadron, its cross section 
for interaction with 
any other ordinary hadron must increase smoothly at higher energies, in much 
the same way as 
the proton-proton or pion-proton cross sections. The underlying reason is the 
increasing role
played by perturbative QCD dynamics and the manifestation of the partonic 
nature of all hadrons. 
Among the existing calculations, no one is strictly valid at $\sqrt{s} 
\simeq 20$ GeV, except 
the one of ref. \cite{hdnnr}, which is designed to work at very high energies 
and which gives, for
the $J/\psi$-nucleon cross section the value $\sigma_{J/\psi -N} \simeq 5$ mb. 
This number can be considered as a guide for $J/\psi - \pi$ cross section in 
the high energy regime. It should, however, be 
pointed out, that the calculation of ref. \cite{hdnnr} is based on a purely 
nonperturbative QCD 
approach. The inclusion of a perturbative contribution will add to the quoted 
value and will 
have a larger weight at higher energies. A similar conclusion was reached in 
\cite{gerland}. 
In the traditional short distance QCD 
approach the cross section grows monotonically \cite{bhp,kha2,lo}. 

As a side-product, the theoretical effort to estimate the charmonium-hadron 
cross section motivated a series of calculations \cite{nosform,nos,bnn}, 
within the framework of QCD sum rules, 
of form factors and coupling constants involving charmed hadrons, that may be 
relevant also to other problems in hadron physics.

In this work we improve the calculation done in ref.~\cite{nnmk} by considering
sum rules based on a three-point function with a pion. We work up to twist-4,
which allows us to study the convergence of the OPE expansion. Since
the method of the QCDSR uses QCD explicitly, we believe that our work 
brings a significant progress to  this important topic.

The paper is organized as follows: in the next section we review the method 
of QCD sum 
rules, giving special emphasis to the QCD side. In section III we present some 
formulas for 
the computation of open charm production amplitudes and in section IV we give 
our numerical results. Finally some concluding remarks are given in section V.

\section{The Method}

In the QCDSR approach \cite{svz,rry}, the short range perturbative QCD is 
extended by an OPE expansion of the correlator, giving a series in inverse 
powers of
the squared momentum with Wilson coefficients. The convergence at low
momentum is improved by using a Borel transform. The coefficients involve 
universal quark and gluon condensates. The quark-based calculation of
a given correlator is equated to the same correlator, calculated using 
hadronic degress of freedom via a dispersion relation, giving sum rules
from which a hadronic quantity can be estimated. 

Let us start with the general vaccum-pion correlation function:
\beqa
\Pi_{\mu34} = \int d^4x~d^4y~e^{-ip_2.y}~e^{ip_3.x}~
\langle 0|T\{j_3(x)j_4(0)j_\mu^\psi(y)\}|\pi(p_1)\rangle \;, 
\label{cor}
\eeqa
with the currents given by  $j_\mu^\psi=\overline{c} \gamma_\mu c$,
$j_3=\overline{u} \Gamma_3 c $ and
$j_4=\overline{c} \Gamma_4d$.
$p_1$, $p_2$, $p_3$ and $p_4$ are the four-momenta of the 
mesons $\pi$, $J/\psi$, $M_3$ and $M_4$ respectively, and $\Gamma_3$ and 
$\Gamma_4$ 
denote specific gamma matrices corresponding to the process envolving the 
mesons $M_3$ and $M_4$. For instance, for the process 
$J/\psi~\pi\rightarrow \bar{D}~D^*$ we will have $\Gamma_3=\gamma_\nu$ and
$\Gamma_4=i\gamma_5$. The advantage of this approach as compared with the 
4-point calculation in ref.~\cite{nnmk}, is that we can consider more terms
in the OPE expansion of the correlation function in Eq.~(\ref{cor}) and,
therefore, check the ``convergence'' of the OPE expansion.

Following ref.~\cite{kl}, we can rewrite Eq.~(\ref{cor}) as:
\beqa
\Pi_{\mu34}=-\int {d^4k\over(2\pi)^4}~Tr[S_{ac}(p_3-k)\gamma_\mu 
S_{cb}(p_3-p_2-k)\Gamma_4 D_{ab}(k,p_1)\Gamma_3]\;,
\label{corqq}
\eeqa 
where  
\beq
S_{ab}(p)=i{\fslash{p}+m_c\over p^2-m_c^2}\delta_{ab}
\eeq
is the free $c$-quark propagator, and $D_{ab}(k,p)$ denotes the
quark-antiquark component with a pion, which can be separated into three 
pieces depending on the Dirac matrices involved \cite{kl,klo}:
\beq
 D_{ab} (k,p) = \delta_{ab} \left [ i\gamma_5 A+
\gamma_\alpha \gamma_5 B^\alpha 
+\gamma_5 \sigma_{\alpha \beta} C^{\alpha \beta}\right]\ ,
\eeq
with $a,\,b$ and $c$ being color indices.
The three invariant functions of $k,p$ are defined by
\beqa
A(k,p)={1\over12}\int d^4x~e^{ik.x}\langle0|\bar{d}(x)i\gamma_5 u(0)
|\pi(p)\rangle \;, 
\nonumber\\
B^\alpha (k,p)={1\over12}\int d^4x~e^{ik.x}\langle0|\bar{d}(x)\gamma^\alpha
\gamma_5 u(0)|\pi(p)\rangle \;, 
\nonumber\\
C^{\alpha \beta} (k,p)=-{1\over24}\int d^4x~e^{ik.x}\langle0|\bar{d}(x)
\sigma^{\alpha\beta}\gamma_5 u(0)|\pi(p)\rangle \;.
\eeqa

Using the soft-pion theorem, PCAC and working at the order 
${\cal O}(p_{\mu} p_{\nu})$ we get up to twist-4 \cite{kl,bbk}:
\beqa
A(k,p)&=&{(2\pi)^4\over12}{\langle {\bar q} q \rangle \over f_\pi}\left[
-2+ip_{\alpha_1}{ \partial\over  i\partial k_{\alpha_1}}
+{1\over3}p_{\alpha_1} p_{\alpha_2}{ \partial\over  i\partial 
k_{\alpha_1}}{ \partial\over  i\partial k_{\alpha_2}}\right]
\delta^{(4)}(k)\;,
\nonumber\\
B_\alpha(k,p)&=&{(2\pi)^4\over12}f_\pi\left[ip_\alpha+{1\over2}
p_\alpha p_{\alpha_1} { \partial\over  i\partial k_{\alpha_1}}
+{i\delta^2\over36}
\biggl(5p_\alpha g_{\alpha_1\alpha_2}
-2p_{\alpha_2} g_{\alpha\alpha_1}\biggr)
{ \partial\over  i\partial 
k_{\alpha_1}}{ \partial\over  i\partial k_{\alpha_2}}\right]
\delta^{(4)}(k)\;,
\nonumber\\
C_{\alpha\beta}(k,p)&=&-{(2\pi)^4\over24}{\langle {\bar q} q \rangle 
\over 3f_\pi}(p_\alpha g_{\beta\alpha_1}-p_\beta g_{\alpha \alpha_1})
\left[i{ \partial\over  i\partial k_{\alpha_1}}
-{p_{\alpha_2}\over2}
{ \partial\over  i\partial 
k_{\alpha_1}}{ \partial\over  i\partial k_{\alpha_2}}\right]
\delta^{(4)}(k)\;,
\label{ABC}
\eeqa
where $\delta^2$ is defined by the matrix element
$\langle 0| {\bar d} g_s {\tilde {\cal G}}^{\alpha\beta}
\gamma_\beta u | \pi(p) \rangle = i \delta^2 f_\pi p^\alpha$,
where ${\tilde {\cal G}}_{\alpha\beta}=\epsilon_{\alpha\beta\sigma\tau}
{\cal G}^{\sigma \tau}/2$
and ${\cal G}_{\alpha \beta} = t^A G_{\alpha \beta}$. 

The additional contributions to the OPE comes from the diagrams 
where one gluon, emitted from the $c$-quark propagator, is combined
with the quark-antiquark component. 
Specifically, the $c$-quark propagator with one gluon being attached
is given by~\cite{rry}
\begin{eqnarray}
-{g_s {\cal G}_{\alpha \beta} \over 2 (k^2 -m^2_c)^2 }
\left [ k_\alpha \gamma_\beta - k_\beta \gamma_\alpha
+ (\fslash{k} + m_c) i \sigma_{\alpha \beta} \right ]\ ,
\end{eqnarray}
Taking the gluon stress tensor into the quark-antiquark component, one
can write down the correlation function into the form
\begin{eqnarray}
\Pi_{\mu34} = -4 \int {d^4 k \over (2\pi)^4}
Tr \bigg[ \bigg(S_{\alpha\beta}(p_3-k)\gamma_\mu S(p_3-p_2-k)
\nonumber \\
+S(p_3-k)\gamma_\mu S_{\alpha\beta}(p_3-p_2-k)\bigg)\Gamma_4 D^{\alpha
\beta}(k,p_1)\Gamma_3\bigg]\;,
\label{corqqg}
\end{eqnarray}
where we have already contracted the color indices, and where we have defined
\beq
S_{\alpha\beta}(k)=-{1\over 2 (k^2 -m^2_c)^2 }
\left [ k_\alpha \gamma_\beta - k_\beta \gamma_\alpha
+ (\fslash{k} + m_c) i \sigma_{\alpha\beta} \right ]\ ,
\eeq
and 
\beq
D^{\alpha\beta}(k,p)= \gamma_5 \sigma_{\rho\lambda} E^{\rho\lambda\alpha
\beta}(k,p)
+\gamma^\tau \epsilon^{\alpha\beta\theta\delta}
F_{\tau\theta\delta}(k,p)\;, 
\eeq
with
\begin{eqnarray}
E^{\rho\lambda\alpha
\beta}(k,p) &=& -{1\over32}\int d^4x~e^{ik.x}
\langle 0 | {\bar d}(x)\gamma_5\sigma^{\rho\lambda} g_s {G}^
{\alpha\beta} u| \pi (p) \rangle \,,
\nonumber \\
F_{\tau\theta\delta}(k,p) &=& {1 \over 32} \int d^4x~e^{ik.x}
\langle 0 | {\bar d}(x) \gamma_\tau g_s
{\tilde G}_{\theta\delta}u | \pi (p) \rangle \,.
\end{eqnarray}

Up to  twist-4 and at order ${\cal O} (p_\mu p_\nu)$, the two functions 
appearing above are given by \cite{kl,bbk}
\begin{eqnarray}
E^{\rho\lambda\alpha\beta} &=& {i\over32}
f_{3\pi} \left(p^\alpha p^\rho g^{\lambda\beta}
-p^\beta p^\rho 
g^{\lambda\alpha}-p^\alpha p^\lambda g^{\rho\beta} +p^\beta p^\lambda 
g^{\rho\alpha}\right) (2\pi)^4\delta^{(4)}(k) 
\nonumber\\
F_{\tau\theta\delta} &=& -{i \delta^2 f_\pi \over 3 \times 32}
(p_\theta g_{\tau\delta} - p_\delta g_{\tau \theta})
(2\pi)^4 \delta^{(4)}(k) \ ,
\label{EF}
\end{eqnarray}
where $f_{3\pi}$ is defined by the vacuum-pion matrix element
$\langle 0| {\bar d} g_s \sigma_{\alpha\beta}\gamma_5{\tilde {\cal G}}^{\alpha
\beta}u | \pi(p) \rangle$ \cite{bbk}.

The phenomenological side of the correlation function, $\Pi_{\mu34}$,
is obtained by the consideration of $J/\psi$, $M_3$ and $M_4$ state 
contribution to the matrix element in Eq.~(\ref{cor}).
The hadronic amplitudes 
are defined by the matrix element:
\beqa
i{\cal{M}}&=&\langle\psi(p_2,\mu)|~M_3(-p_3,\nu)~M_4(-p_4,\rho)~\pi(p_1)
\rangle
\nonumber\\
&=&i~{\cal{M}}_{\mu34}(p_1,p_2,p_3,p_4)~\epsilon_2^{\mu}
f_3^{*\nu}f_4^{*\rho}\;,
\eeqa
where $f_i^{*\alpha}=\epsilon_i^{*\alpha}$ for the $D^*$ meson and $f_i^
{*\alpha}=1$ for the $D$ meson.

The phenomenological side of the sum rule can be written as (for the
part of the hadronic amplitude that will contribute to the cross section)
\cite{nnmk}:
\beq
\Pi_{\mu34}^{phen}=-{\mpsi f_\psi\lambda_3\lambda_4
~{\cal{M}}_{\mu34}\over
(p_2^2-\mpsi^2)(p_3^2-m_3^2)(p_4^2-m_4^2)} + \mbox{h. r.}\; ,
\label{phendds}
\eeq
where h.~r. means higher resonances, and where $\lambda_i$ is related with
the corresponding meson decay constant:
$\langle D|j_D|0\rangle=-\lambda_D={\md^2 f_D/ m_c}$
and $\langle 0|j_\alpha|D^*\rangle=\lambda_{D^*}\epsilon_\alpha=
\mds f_{D^*}\epsilon_\alpha$.

\section{Hadronic Amplitudes for $J/\psi~\pi\rightarrow$ Open Charm}

The hadronic amplitudes can be written in terms of many different structures.
In terms of the structures that will contribute to the cross section we
can write
\begin{itemize} \item for the process $J/\psi~\pi\rightarrow \bar{D}~D^*$:
\beqa
{\cal{M}}_{\mu\nu}=\Lambda_1^{DD^*}p_{1\mu}p_{1\nu} + \Lambda_2^{DD^*}
p_{1\mu}p_{2\nu} +\Lambda_3^{DD^*}p_{1\nu}p_{3\mu} + \Lambda_4^{DD^*} 
g_{\mu\nu}+ 
\Lambda_5^{DD^*}p_{2\nu}p_{3\mu}\;,
\label{strudds}
\eeqa
\item for the process $J/\psi~\pi\rightarrow \bar{D}~D$:
\beq
{\cal{M}}_{\mu}=\Lambda_{DD}~\epsilon_{\mu\alpha\beta\sigma}p_{1}^\alpha 
 p_3^\beta p_4^\sigma\,,
\label{strudd}
\eeq
\item for the process $J/\psi~\pi\rightarrow \bar{D}^*~D^*$:
\beqa
&&{\cal{M}}_{\mu\nu\rho}=
\Lambda_1^{D^*D^*}~H_{\mu\nu\rho}+ \Lambda_2^{D^*D^*}~J_{\mu\nu\rho}
+\Lambda_3^{D^*D^*}g_{\nu\rho}\epsilon_{\mu\alpha\beta\gamma}p_{1}^\alpha 
 p_2^\beta p_3^\gamma + \Lambda_4^{D^*D^*}\epsilon_{\nu\rho\alpha\beta}
p_{3\mu} p_{1}^\alpha p_3^\beta 
\nonumber\\
&+& \Lambda_5^{D^*D^*}
\epsilon_{\nu\rho\alpha\beta}p_{3\mu} p_{1}^\alpha  p_2^\beta 
+\Lambda_6^{D^*D^*}\epsilon_{\mu\nu\alpha\beta}p_{3\rho}p_{1}^\alpha 
 p_2^\beta 
+ \Lambda_7^{D^*D^*}\epsilon_{\mu\nu\alpha\beta}
p_{1\rho} p_{1}^\alpha 
 p_2^\beta 
\nonumber\\
&+& \Lambda_8^{D^*D^*}\epsilon_{\nu\rho\alpha\beta}p_{1\mu} 
p_{1}^\alpha  p_4^\beta 
+\Lambda_9^{D^*D^*}\epsilon_{\mu\nu\rho\alpha}p_{1}^\alpha 
+ \Lambda_{10}^{D^*D^*}\epsilon_{\nu\rho\alpha\beta}p_{1\mu} 
p_{1}^\alpha  p_3^\beta 
+\Lambda_{11}^{D^*D^*}\epsilon_{\mu\nu\rho\alpha}
p_{2}^\alpha 
\nonumber\\
&+&\Lambda_{12}^{D^*D^*}\epsilon_{\mu\nu\alpha\beta}
p_{1\rho} p_{1}^\alpha  p_3^\beta + \Lambda_{13}^{D^*D^*}
\epsilon_{\mu\nu\alpha\beta}p_{3\rho} p_{1}^\alpha  p_3^\beta \;.
\label{strudsds}
\eeqa
\end{itemize} 
with
$H_{\mu\nu\rho}=(\epsilon_{\nu\alpha\beta\gamma}g_{\mu\rho}-\epsilon_{\rho
\alpha\beta\gamma}g_{\mu\nu})p_{1}^\alpha p_2^\beta p_3^\gamma + 
\epsilon_{\mu\rho\alpha\beta}p_{2\nu} p_1^\alpha p_2^\beta$ and
$J_{\mu\nu\rho}=(\epsilon_{\nu\rho\alpha\beta}p_{1\mu}+\epsilon_{\mu\rho
\alpha\beta}p_{1\nu}
+\epsilon_{\mu\nu\alpha\beta}p_{1\rho})p_{2}^\alpha p_3^\beta +
\epsilon_{\mu\nu\alpha\beta}p_{2\rho}p_{1}^\alpha p_3^\beta$.

In Eqs.~(\ref{strudds}), (\ref{strudd}) and (\ref{strudsds}), $\Lambda_i$ 
are the parameters that we will evaluate from the sum rules. In principle
all the independent structures appearing in $H_{\mu\nu\rho}$ and 
$J_{\mu\nu\rho}$ would have independent parameters $\Lambda_i$. However,
since in our approach we get exactly the same sum rules for all of them,
we decided to group them with the same parameters.

Inserting the results in Eqs.~(\ref{ABC}) and (\ref{EF}) into 
Eqs.~(\ref{corqq}) and (\ref{corqqg}) we can write a sum rule for each of 
the structures appearing in Eqs.~(\ref{strudds}), (\ref{strudd}) and 
(\ref{strudsds}). To improve the matching between
the phenomenological and theoretical sides we follow the usual procedure and
make a single Borel transformation to  all the external momenta taken to be 
equal: $-p_2^2=-p_3^2=-p_4^2=P^2\rightarrow M^2$. The problem of doing a 
single Borel transformation is the fact that terms 
associated with the pole-continuum  transitions are not
suppressed~\cite{io2}. In the present case we have two kinds of these 
transitions: double pole-continuum and single pole-continuum. In the limit
of similar meson masses it is easy to show that the Borel behaviour of
the three-pole,  double pole-continuum and single pole-continuum
contributions are $e^{-m_M^2/M^2}/M^4,\,e^{-m_M^2/M^2}/M^2$ and
$e^{-m_M^2/M^2}$ respectively. Therefore, we can single out the three-pole
contribution from the others by introducing two parameters in the 
phenomenological side of the sum rule, which will account for the 
double pole-continuum and single pole-continuum contributions. The 
expressions for all 19 sum rules are given in the Appendices A, B and C.

\section{Results and Discussion}

The parameter values used in all calculations are  
$m_c=1.37\,\GeV$, $m_\pi=140\,\MeV$, $m_D=1.87\,\GeV$, $m_{D^*}=2.01\,
\GeV$, $\mpsi=3.097\,\GeV$, $f_\pi=131.5\,\MeV$, 
$\langle\overline{q}q\rangle=-(0.23)^3\,\GeV^3$, $m_0^2=0.8\,\GeV^2$,
$\delta^2=0.2\,\GeV^2$, $f_{3\pi}=0.0035\,\GeV^2$ \cite{bbk}.
For the charmed mesons decay constants we use the values from \cite{bbk}
for $f_D$ and $f_{D^*}$ and the experimental value for $f_\psi$:
\beq
f_\psi=270\,\MeV,\;\;\;\;\;
f_D=170\,\MeV,\;\;\;\;f_{D^*}=240\,\MeV.
\label{num}
\eeq

In ref.~\cite{dlnn} we have analyzed the  sum rule for the process
$J/\psi~\pi\rightarrow \bar{D}~D$. Here we choose to show the sum rule
for $\Lambda_1^{D*D*}$ in Eq.~(\ref{c1}), as an example of the sum rules
for the process  $J/\psi~\pi\rightarrow \bar{D}^*~D^*$.

\begin{figure}[htb]
\centerline{\psfig{figure=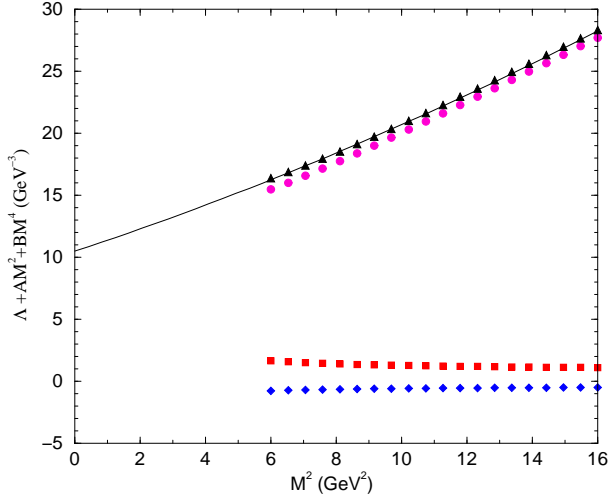,width=8cm,angle=0}}
\protect\caption{Sum rule for $\Lambda_1^{D^*D^*}$ related to the process 
$J/\psi~\pi\rightarrow \bar{D}^*~D^*$
as a function of the Borel mass. The dots, squares and diamonds 
give the twist-2, 3 and 4 contributions to the sum rule. The triangles
give the result from  Eq.~(\ref{c1}). The solid line give the fit
to the QCDSR results.}
\label{fig1}
\end{figure}

In Fig.~1 we show the QCD sum rule results for 
$\Lambda_1^{D^*D^*}+A_1^{D^*D^*}M^2+B_1^{D^*D^*}M^4$ as a function of $M^2$. 
The dots, squares and 
diamonds give the twist-2, 3 and 4 contributions respectively. The triangles
give the final QCDSR results. We see that the twist-3 and 4 contributions
are small as compared with the twist-2 contribution, following the same
behaviour as the sum rule for the process $J/\psi~\pi\rightarrow \bar{D}~D$
given in \cite{dlnn}. In general all the other sum rules are 
similar and contain  twist-2, twist-3 and twist-4 contributions corresponding
to the first, second, and third terms inside the brackets in the right hand 
side of Eq.~(\ref{c1}). Only the sum rules for $\Lambda_{10}^{D^*D^*}$ up
to $\Lambda_{13}^{D^*D^*}$, $\Lambda_{4}^{DD^*}$ and $\Lambda_{5}^{DD^*}$
do not get the leading twist contribution, and will be neglected
in the evaluation of the cross section. 
It is also interesting to notice that if we consider
only the leading twist contributions we recover the sum rules obtained in 
ref.~\cite{nnmk}. The triangles in Fig.~1
follow almost a straight line in the Borel region $6\leq M^2\leq16\,\GeV^2$.
This show that the single pole-continuum transitions contribution is small.
The value of the amplitude $\Lambda_1^{D^*D^*}$ is obtained by the 
extrapolation of
the fit to $M^2=0$~\cite{bnn,nos,io2}. Fitting the QCD sum rule results
to a quadratic form we get 
\beq
\Lambda_1^{D^*D^*}\simeq10.5\GeV^{-3}.
\label{ampli}
\eeq
Since we worked in the soft pion limit, $\Lambda_1^{D^*D^*}$, as well as all 
other $\Lambda$, is just a number. All 
particle momenta dependence of the amplitudes is contained 
in the Dirac structure. 

In obtaining the results shown in Fig.~1 we have used the numerical
values for the meson decay constants given in Eq.~(\ref{num}). However, it
is also possible to use the respective sum rules, as done in \cite{nnmk}. 
The two-point sum rules for the meson decay constants are given in the 
appendix D. The behaviour of the results for the hadronic amplitudes
does not change significantly if we use the two-point sum rules for the
meson decay constants instead of the numerical values, leading only to a 
small change in the value of the amplitudes. In Fig.~2 we show, for a 
comparison, both results in the case of $\Lambda_1^{D^*D^*}$.
\begin{figure}[htb]
\centerline{\psfig{figure=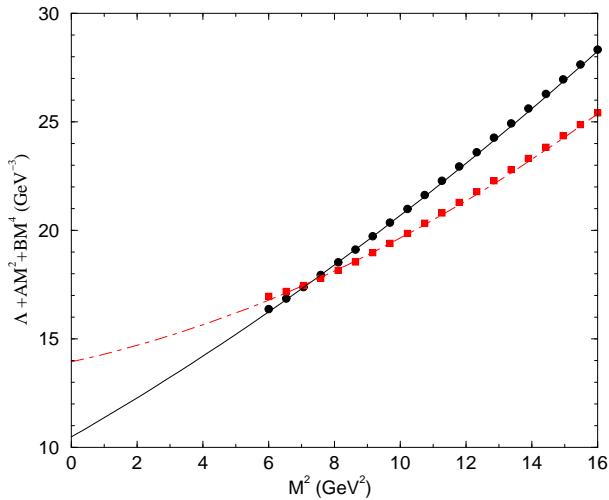,width=8cm,angle=0}}
\protect\caption{Sum rule for $\Lambda_1^{D^*D^*}$ related to the process 
$J/\psi~\pi\rightarrow \bar{D}^*~D^*$
as a function of the Borel mass. The dots and squares give the results from  
Eq.~(\ref{c1}) when using respectively numerical values or the two-point 
sum rules for the meson decay constants. The solid and dot-dashed lines give 
the fits to the QCDSR results.}
\label{fig2}
\end{figure}

Using the respective sum rules for the meson decay constants we get
\beq
\Lambda_1^{D^*D^*}\simeq13.9\GeV^{-3}.
\label{ampli2}
\eeq
We will use these two procedures to estimate
the errors in our calculation. It is important to mention that our results  
agree completely with the value obtained in \cite{nnmk}.

The results for all other sum rules show a similar behaviour 
and the amplitudes can be extracted by the extrapolation of the fit to 
$M^2=0$. The QCDSR results, evaluated using the numerical values
for the meson decay constants, as well as the quadratic fits for the 
amplitudes associated with the process $J/\psi~\pi\rightarrow\bar{D}~D^*$ 
are shown in Fig.~3.
\begin{figure}[htb]
\centerline{\psfig{figure=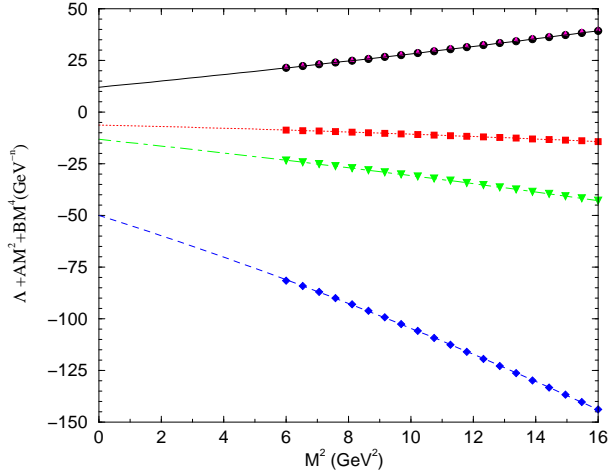,width=8cm,angle=0}}
\protect\caption{Sum rules for $\Lambda_i^{DD^*}$ related to the process 
$J/\psi~\pi\rightarrow \bar{D}~D^*$
as a function of the Borel mass. The dots, squares, diamonds, triangles up 
and triangles down give the results from the QCDSR for $\Lambda_1^{DD^*}$
up to $\Lambda_5^{DD^*}$ respectively. The QCDSR results for $\Lambda_1^{DD^*}$
and $\Lambda_4^{DD^*}$ can not be distinguished at the scale of the figure.
The solid, dotted, dashed, long-dashed 
and dot-dashed lines give the fits to the QCDSR results.}
\label{fig3}
\end{figure}

The values for the parameters associated with the process $J/\psi~\pi
\rightarrow\bar{D}~D^*$ are given in Table I.
\begin{table}[htb]
\caption{\label{tab:table1}The best fitted values for the parameters
associated with the process $J/\psi~\pi\rightarrow\bar{D}~D^*$.}
\begin{ruledtabular}
\begin{tabular}{ccccc}
$\Lambda_1^{DD^*}$ & $\Lambda_2^{DD^*}$ & $\Lambda_3^{DD^*}$ & 
$\Lambda_4^{DD^*}$ & $\Lambda_5^{DD^*}$ \\
\hline
$14\pm2 \GeV^{-2}$  & $-7.2\pm0.9$ GeV$^{-2}$ & $-58\pm8 \GeV^{-2}$ & 
$14.6\pm2.2$ & $-15.6\pm2.2 \GeV^{-2}$\\
\end{tabular}
\end{ruledtabular}
\end{table}

For the process $J/\psi~\pi\rightarrow\bar{D}~D$ we have only one parameter
which is given by \cite{dlnn}
\beq
\Lambda_{DD}=13.2\pm1.8\GeV^{-3},
\label{amplidd}
\eeq
and the 13 parameters associated with the process $J/\psi~\pi
\rightarrow\bar{D}^*~D^*$ are given in Table II.
\begin{table}[htb]
\caption{\label{tab:table2}The best fitted values for the parameters
associated with the process $J/\psi~\pi\rightarrow\bar{D}^*~D^*$.}
\begin{ruledtabular}
\begin{tabular}{ccccc}
$\Lambda_1^{D^*D^*}$ & $\Lambda_2^{D^*D^*}$ & $\Lambda_3^{D^*D^*}$ & 
$\Lambda_4^{D^*D^*}$ & $\Lambda_5^{D^*D^*}$ \\
\hline
$12.2\pm1.7 \GeV^{-3}$  & $-12.8\pm1.8$ GeV$^{-3}$ & $12.5\pm1.7 \GeV^{-3}$ & 
$-24.6\pm3.4 \GeV^{-3}$ & $9.8\pm1.6 \GeV^{-3}$\\
\hline
\hline
$\Lambda_6^{D^*D^*}$ & $\Lambda_7^{D^*D^*}$ & $\Lambda_8^{D^*D^*}$ & 
$\Lambda_9^{D^*D^*}$ & $\Lambda_{10}^{D^*D^*}$ \\
\hline
$9.7\pm1.6 \GeV^{-3}$  & $-13.0\pm1.8$ GeV$^{-3}$ & $-13.8\pm1.8 \GeV^{-3}$ & 
$-5.4\pm0.9 \GeV^{-1}$ & $2.5\pm0.2 \GeV^{-3}$\\
\hline
\hline
$\Lambda_{11}^{D^*D^*}(\GeV^{-1})$ & $\Lambda_{12}^{D^*D^*}$ & 
$\Lambda_{13}^{D^*D^*}$  &&\\
\hline
$(-5.5\pm0.5)10^{-3}$  &$-0.022\pm0.002$ GeV$^{-3}$ 
& $0.53\pm0.03 \GeV^{-3}$ &&  \\
\end{tabular}
\end{ruledtabular}
\end{table}
The errors in all parameters were estimated by the evaluation of the sum 
rules using the numerical
values and the two-point QCDSR for the meson decay constants.

Having the QCD sum rule results for the amplitudes of the three processes
$J/\psi~\pi\rightarrow \bar{D}~D^*,~\bar{D}~D,~\bar{D}^*~D^*$, given in 
Eqs.~(\ref{strudds}), (\ref{strudd}) and (\ref{strudsds})
we can evaluate  the cross sections.
After including isospin factors, the differential cross section for 
the $J/\psi-\pi$ dissociation is given by
\beq
{d\sigma\over dt}={1\over 96\pi s{\bf p}_{i,cm}^2}~\sum_{spin}|{\cal M}|^2~,
\label{sig}
\eeq
where ${\bf p}_{i,cm}$ is the three-momentum of $p_1$ (or $p_2$) in the center
of mass frame (with $p_1~(p_2)$ being the four-momentum of the 
$\pi~(J/\psi)$):
\beq
{\bf p}_{i,cm}^2={\lambda(s,\mpi^2,\mpsi^2)\over4s}~,
\eeq
with $\lambda(x,y,z)=x^2+y^2+z^2-2xy-2xz-2yz$,
$s=(p_1+p_2)^2$, $t=(p_1-p_3)^2$.

In Eq.~(\ref{sig}), the sum over the spins of the amplitude squared is given 
by
\beq
\sum_{spin}|{\cal M}|^2~=~
{\cal M}_{\mu\nu}{\cal M}_{\mu'\nu'}^*\left(g^{\mu\mu'}-{p_2^\mu
p_2^{\mu'}\over\mpsi^2}\right)\left(g^{\nu\nu'}-{p_3^\nu
p_3^{\nu'}\over\mds^2}\right)\;,
\label{m2ds}
\eeq
for $J/\psi~\pi\rightarrow \bar{D}~D^*$, with $p_3~(p_4)$ being the 
four-momentum of $D^*~(D)$.
\beq
\sum_{spin}|{\cal M}|^2~=~
{\cal M}_{\mu}{\cal M}_{\mu'}^*\left(g^{\mu\mu'}-{p_2^\mu
p_2^{\mu'}\over\mpsi^2}\right)\;,
\label{m2dd}
\eeq
for $J/\psi~\pi\rightarrow \bar{D}~D$, and 
\beq
\sum_{spin}|{\cal M}|^2~=~
{\cal M}_{\mu\nu\alpha}{\cal M}_{\mu'\nu'\alpha'}^*\left(g^{\mu\mu'}-{p_2^\mu
p_2^{\mu'}\over\mpsi^2}\right)\left(g^{\nu\nu'}-{p_3^\nu
p_3^{\nu'}\over\mds^2}\right)\left(g^{\alpha\alpha'}-{p_4^\alpha
p_4^{\alpha'}\over\mds^2}\right)\;,
\label{m2ss}
\eeq
for $J/\psi~\pi\rightarrow \bar{D}^*~D^*$.

\begin{figure}[htb]
\centerline{\psfig{figure=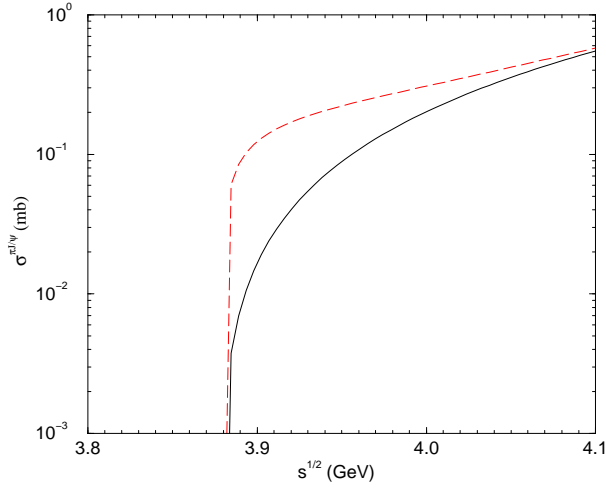,width=8cm,angle=0}}
\protect\caption{$J/\psi~\pi\rightarrow \bar{D}~D^*+\bar{D}^*~D$ cross 
section. The solid and dashed lines give the results which respect and break 
chiral symmetry respectively.} 
\label{fig4}
\end{figure}
The structures multiplying $\Lambda_4^{DD^*}$ and  
$\Lambda_5^{DD^*}$ in Eq.~(\ref{strudds}), and  $\Lambda_{11}^{D^*D^*}$ in 
Eq.~(\ref{strudsds}) break chiral symmetry \cite{nnr}. To evaluate the effect 
of breaking chiral symmetry in the process $J/\psi~\pi\rightarrow \bar{D}~D^*
+\bar{D}^*~D$ we show, in Fig.~4, the cross section calculated using all 
structures in 
Eq.~(\ref{strudds}) (dashed line) and neglecting $\Lambda_4^{DD^*}$ and  
$\Lambda_5^{DD^*}$ (solid line).

From Fig.~4 we see that the cross section obtained with the amplitude that
breaks chiral symmetry grows very fast near the threshold. Since this is the 
energy region where this kind of process is probable more likely to happen,
it is very important to use models that respect chiral symmetry when evaluating
the $J/\psi-\pi$ cross section. 
\begin{figure}[htb]
\centerline{\psfig{figure=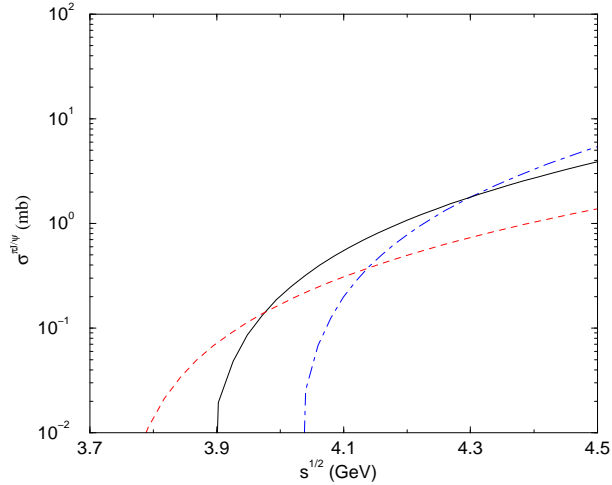,width=8cm,angle=0}}
\protect\caption{$J/\psi-\pi$ dissociation cross sections for the 
processes $J/\psi~\pi\rightarrow\bar{D}~D^* + D~\bar{D}^*$ (solid line),
$J/\psi~\pi\rightarrow\bar{D}~D$ (dashed line) and
$J/\psi~\pi\rightarrow\bar{D}^*~D^*$ (dot-dashed line).} 
\label{fig5}
\end{figure}
As mentioned before, the sum rules for $\Lambda_4^{DD^*}$,
$\Lambda_5^{DD^*}$ and $\Lambda_{10}^{D^*D^*}$ up to $\Lambda_{13}^{D^*D^*}$,
do not get the leading order contribution and will be neglected when
evaluating the cross sections. It is important to keep in mind that,
since our sum rules were derived in the limit $p_{1}\rightarrow0$,
we can not extend our results to large values of $\sqrt{s}$. For this reason
we will limit our calculation to $\sqrt{s}\leq4.5 \GeV$.

In Fig.~5 we show separately the 
contributions for each one of the process.
Our first conclusion is that our results show that, for  values
of $\sqrt{s}$ far from the $J/\psi~\pi\rightarrow \bar{D}^{*}~{D}^*$
threshold,  $\sigma_{J/\psi\pi\rightarrow \bar{D}^{*}{D}^*} \, \geq \,
\sigma_{J/\psi\pi\rightarrow \bar{D}{D}^*+D\bar{D}^*} \, \geq \,
\sigma_{J/\psi\pi\rightarrow \bar{D}{D}}$, in agreement with the model 
calculations presented in \cite{osl} but in disagreement with 
the results obtained with the nonrelativistic quark model of \cite{wongs}, 
which show that the state $\bar{D}^*D$ has a larger production cross
section than $\bar{D}^{*}{D}^*$. Furthermore, our curves indicate that 
the cross section grows monotonically with the c.m.s. energy but not as fast, 
near the thresholds, as it does in the calculations in
Refs.~\cite{osl,haglin,linko,ikbb,nnr}. Again, this behavior is in
opposition to~\cite{wongs}, where a peak just after the threshold 
followed by continuous decrease in the cross section was found.

\begin{figure}[htb]
\centerline{\psfig{figure=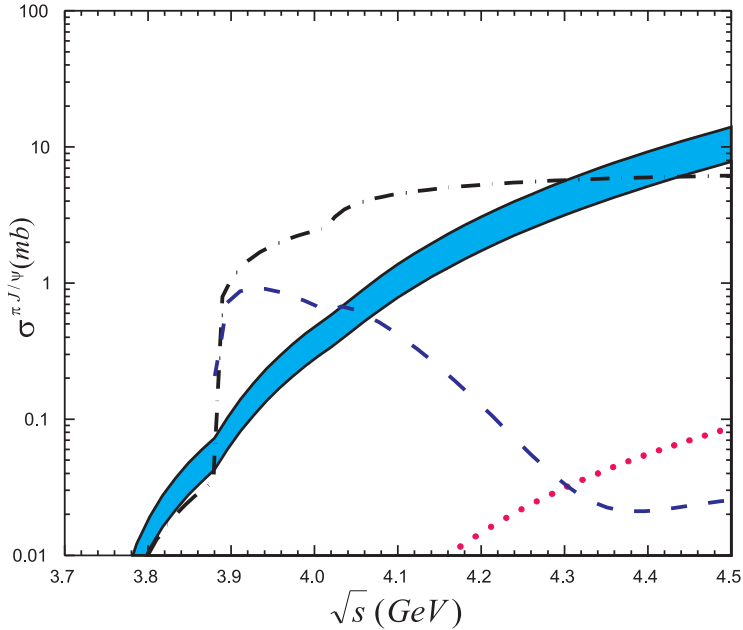,width=10cm,angle=0}}
\protect\caption{$J/\psi-\pi$ dissociation cross sections  from meson
exchange model \cite{osl} (dot-dashed line), quark exchange model \cite{wongs}
(dashed line), short distance QCD \cite{kha2,lo} (dotted line) and QCD sum 
rules (solid lines). The shaded area
give an evaluation of the uncertainties in our calculation.} 
\label{fig6}
\end{figure}

In Fig.~6 we show, for comparison, our result for  the total cross section 
for the $J/\psi~\pi$ dissociation (solid lines) and the results from meson 
exchange model \cite{osl} obtained with a cut-off $\Lambda=1\GeV$
(dot-dashed line), quark exchange model \cite{wongs}
(dashed line) and short distance QCD \cite{kha2,lo} (dotted line). The shaded 
area in our results give an evaluation of the 
uncertainties in our calculation obtained with the two procedures described 
above. It is very interesting to notice that bellow the $DD^*$ threshold, our
result and the results from meson exchange and quark exchange models are
in a very good agreement. However, as soon as the $DD^*$ channel is open
the cross section obtained with the meson exchange and quark exchange models
show a very fast grown, as a function of $\sqrt{s}$, as compared with our
result. As discussed above, this is due to the fact that chiral
symmetry is broken in these two model calculations.

The momentum distribution of thermal pions in a hadron gas depends on the 
effective temperature $T$ with an approximate Bose-Einsten distribution.
Therefore, in a hadron gas, pions collide with the $J/\psi$ at different 
energies, and the relevant quantity is not the value of the cross section
at a given energy, but  the thermal average of the cross section.
The thermal average of the cross section is defined by the product
of the dissociation cross section and the relative velocity 
of initial state particles, averaged
over the energies of the pions: $\langle\sigma^{\pi J/\psi} v\rangle$, and
is given by \cite{linko}
\beq
\langle\sigma^{\pi J/\psi} v\rangle={\int_{z_0}^\infty dz[z^2-(\alpha_1+
\alpha_2)^2][z^2-(\alpha_1-\alpha_2)^2]K_1(z)\sigma^{\pi J/\psi}(s=z^2T^2)\over
4\alpha_1^2K_2(\alpha_1)\alpha_2^2K_2(\alpha_2)}\,,
\eeq
where $\alpha_i=m_i/T~(i=1$ to $4)$, $z_0=$ max$(\alpha_1+\alpha_2,\alpha_3+
\alpha_4)$ and $K_i$ is the modified Bessel function.

\begin{figure}[htb]
\centerline{\psfig{figure=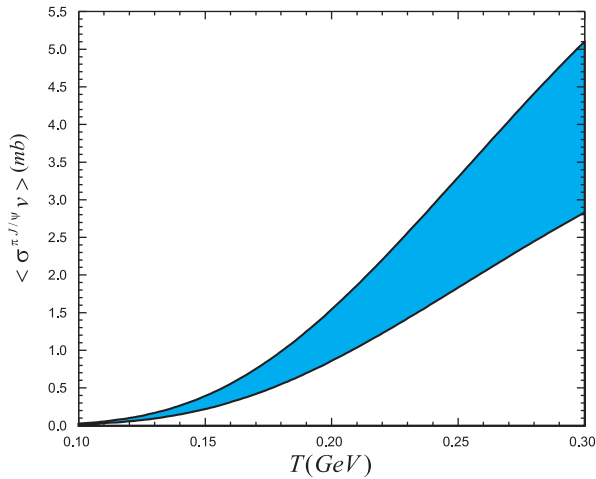,width=8cm,angle=0}}
\protect\caption{Thermal average of $J/\psi$ dissociation cross section by 
pions as a function of temperature $T$. The shaded area
give an evaluation of the uncertainties in our calculation.} 
\label{fig7}
\end{figure}

As shown in Fig.~7, $\langle\sigma^{\pi J/\psi} v\rangle$ increases with the
temperature. Since the $J/\psi$ dissociation by a pion requires energetic 
pions to overcome the energy threshold, it has a small thermal average
at low temperatures. The magnitude of our thermal average cross section
is of the same order as the meson exchange model calculation in 
ref.~\cite{linko} with a cut-off $\Lambda=1\GeV$. The shaded area in Fig.~7
give an evaluation of the uncertainties in our calculation due to the two 
procedures used to extract the hadronic amplitudes.

\section{Summary and Conclusions}

We have evaluated the hadronic amplitudes for the  $J/\psi$ 
dissociation by pions using the QCD sum rules based on a three-point function 
using vaccum-pion correlation functions. We have considered the OPE 
expansion up to twist-4 and we have worked in the soft-pion limit. 
Our work improves  the former QCDSR calculation, done with a four-point 
function at the pion pole \cite{nnmk}, since we have included more terms in
the OPE expansion. We have shown that the twist-3 and twist-4 contributions
to the sum rules are small when compared with the leading order contribution,
showing a good ``convergence'' of the OPE expansion. We have checked that, 
taking the appropriated limit, we recover the previous result of \cite{nnmk}. 

From the theoretical point of view, the use of QCDSR in this problem was 
responsible 
for real progress, being a step beyond models and beyond the previous leading 
twist calculations \cite{bhp,kha2,lo,arleo,nnmk}. 
This is specially true in the low energy region, close to the open charm 
production threshold. At higher energies our
treatment is less reliable due to the approximations employed.

Although a more sophisticated  analysis of our uncertainties is still to be 
done, the shaded area in Fig. 6 shows 
that we can make some unambiguous statement concerning the behavior of
$\sigma_{\pi J/\psi}$ with the energy $\sqrt{s}$. 
Our  cross section grows monotonically with the c.m.s. energy 
but not as fast, near the thresholds, as it does in the calculations 
using meson exchange models \cite{osl,haglin,linko,ikbb,nnr}. We have also 
shown the importance of respecting chiral symmetry, since the increase of
the cross section near the threshold is strongly intensified when chiral 
symmetry is broken. 
In other words, our results suggest that, {\it using meson exchange models is 
perfectly acceptable, provided that they include form factors and  that they  
respect chiral symmetry}. With these precautions, they can be a good tool to 
make predictions at somewhat higher energies.

We have also evaluated the
thermal average of the $J/\psi-\pi$ dissociation cross section. It increases
with the temperature and at $T=150\,\MeV$ we get
$\langle\sigma^{\pi J/\psi} v\rangle\sim0.2-0.4$ mb which is compatible with 
the values presented in Fig. 5 of ref. \cite{linko}, i.e., in a meson 
exchange model with monopole form  
factor with cut-off $\Lambda = 1 $ GeV. The use of this information will 
reduce the uncertainties
in the calculations of the hadronic lifetimes of $J/\psi$, which are needed 
in simulations like those of ref. \cite{rapp}.

\section*{Acknowledgments}

We are grateful to J. H\"ufner for fruitful discussions. We thank Yongseok
Oh for providing the data to construct fig.~6. M.N. would 
like to thank the hospitality and financial support from the Yonsei University
during her stay in Korea. S.H.L and H.K. were supported by Korean Research 
Foundation Grant (KRF-2002-015-CP0074). This work was supported by CNPq and 
FAPESP-Brazil. 

\appendix\section{Sum rules for the process $J/\psi~\pi\rightarrow 
\bar{D}~D^*$}
Using $\Gamma_3=\gamma_\nu$ and $\Gamma_4=i\gamma_5$ in Eqs.~(\ref{corqq}) 
and (\ref{corqqg}), we obtain the following sum rules for the
structures in Eq.~(\ref{strudds}):
\beqa
\Lambda_1^{DD^*}+A_1^{DD^*}M^2+B_1^{DD^*}M^4&=&{1\over~C_{DD^*}~f_{DD^*}(M^2)}
\left[-f_\pi m_c\right.
\nonumber\\
&+&f_{3\pi}\Biggl(3+\left.{m_c^2\over2M^2}\Biggr)\right]
{e^{-m_c^2/M^2}\over M^2}\,,
\eeqa
\beqa
\Lambda_2^{DD^*}+A_2^{DD^*}M^2+B_2^{DD^*}M^4&=&{1\over~C_{DD^*}~f_{DD^*}(M^2)}
\left[f_\pi m_c+{4\langle\bar{q}q\rangle\over
3f_\pi}\right.
\nonumber\\
-{m_c\over6M^2}\Biggl({4\langle\bar{q}q\rangle m_c\over
f_\pi}&+&\left.{\delta^2f_\pi\over3}\left(19+{5m_c^2\over M^2}\right)\Biggr)
\right]{e^{-m_c^2/M^2}\over M^2}\,,
\eeqa
\beqa
\Lambda_3^{DD^*}+A_3^{DD^*}M^2+B_3^{DD^*}M^4&=&{1\over~C_{DD^*}~f_{DD^*}(M^2)}
\left[2f_\pi m_c-{8\langle\bar{q}q\rangle\over
3f_\pi}\right.
\nonumber\\
-{m_c\over6M^2}\Biggl({8\langle\bar{q}q\rangle m_c\over
f_\pi}&+&\left.{\delta^2f_\pi\over3}\left(-2+{10m_c^2\over M^2}\right)
\Biggr)\right]{e^{-m_c^2/M^2}\over M^2}\,,
\eeqa
\beqa
\Lambda_4^{DD^*}+A_4^{DD^*}M^2+B_4^{DD^*}M^4&=&{m_c^2\over~C_{DD^*}~f_{DD^*}
(M^2)}
{\langle\bar{q}q\rangle\over f_\pi}{e^{-m_c^2/M^2}\over M^2}\,,
\eeqa
\beqa
\Lambda_5^{DD^*}+A_5^{DD^*}M^2+B_5^{DD^*}M^4&=&{-2\over~C_{DD^*}~f_{DD^*}(M^2)}
{\langle\bar{q}q\rangle\over f_\pi}{e^{-m_c^2/M^2}\over M^2}\,,
\eeqa
where
\beqa
f_{DD^*}(M^2)= {1\over\mpsi^2-\mds^2}\Biggl[{e^{-\md^2/M^2}-
e^{-\mpsi^2/M^2} \over\mpsi^2-\md^2}~- {e^{-\md^2/M^2}-
e^{-\mds^2/M^2} \over\mds^2-\md^2}\Biggr]\,,
\eeqa
and
\beq
C_{DD^*}={m_c\over\md^2\mds\mpsi f_Df_{D^*}f_\psi}\,.
\eeq
$A_i^{DD^*}$ and $B_i^{DD^*}$ are the parameters introduced to account for
double pole-continuum and single pole-continuum transitions respectively.

\section{Sum rules for the process $J/\psi~\pi\rightarrow \bar{D}~D$}

Using $\Gamma_3=i\gamma_5$ and $\Gamma_4=i\gamma_5$ in Eqs.~(\ref{corqq}) 
and (\ref{corqqg}), we obtain the following sum rule for the
structure in Eq.~(\ref{strudd}) \cite{dlnn}:
\beqa
{\Lambda_{DD}+A_{DD}M^2+B_{DD}M^4\over\mpsi^2-\md^2}\Biggl[{e^{-\md^2/
M^2}\over M^2}- {e^{-\md^2/M^2}-
e^{-\mpsi^2/M^2} \over\mpsi^2-\md^2}\Biggr]
\nonumber\\
= {m_c^2\over\md^4\mpsi f_{D}^2f_\psi}
{e^{-m_c^2/M^2}\over M^2}  \Biggl[f_\pi-{2m_c\langle\bar{q}q\rangle\over
3f_\pi M^2}
-{f_\pi\delta^2\over18M^2}\Biggl(17+{5m_c^2\over M^2}\Biggr)\Biggr].
\label{sr}
\eeqa

\section{Sum rules for the process $J/\psi~\pi\rightarrow 
\bar{D}^*~D^*$}

Using $\Gamma_3=\gamma_\nu$ and $\Gamma_4=\gamma_\rho$ in Eqs.~(\ref{corqq}) 
and (\ref{corqqg}), we obtain the following sum rules for the
structures in Eq.~(\ref{strudsds}):
\beqa
\Lambda_1^{D^*D^*}+A_1^{D^*D^*}M^2+B_1^{D^*D^*}M^4&=&{1\over
C_{D^*D^*}~f_{D^*D^*}(M^2)}
\Biggl[-f_\pi+{2m_c\langle\bar{q}q\rangle\over
3f_\pi M^2}
\nonumber\\
&+&{f_\pi\delta^2\over6M^2}\Biggl({25\over3}+{5m_c^2\over3M^2}\Biggr)\Biggr]
{e^{-m_c^2/M^2}\over M^2}\,,
\label{c1}
\eeqa
\beqa
\Lambda_2^{D^*D^*}+A_2^{D^*D^*}M^2+B_2^{D^*D^*}M^4&=&{1\over
C_{D^*D^*}~f_{D^*D^*}(M^2)}
\Biggl[f_\pi-{2m_c\langle\bar{q}q\rangle\over
3f_\pi M^2}
\nonumber\\
&-&{f_\pi\delta^2\over6M^2}\Biggl({13\over3}+{5m_c^2\over3M^2}\Biggr)\Biggr]
{e^{-m_c^2/M^2}\over M^2}\,,
\eeqa
\beqa
\Lambda_3^{D^*D^*}+A_3^{D^*D^*}M^2+B_3^{D^*D^*}M^4&=&{1\over
C_{D^*D^*}~f_{D^*D^*}(M^2)}
\Biggl[-f_\pi+{2m_c\langle\bar{q}q\rangle\over
3f_\pi M^2}
\nonumber\\
&+&{f_\pi\delta^2\over6M^2}\Biggl({19\over3}+{5m_c^2\over3M^2}\Biggr)\Biggr]
{e^{-m_c^2/M^2}\over M^2}\,,
\eeqa
\beqa
\Lambda_4^{D^*D^*}+A_4^{D^*D^*}M^2+B_4^{D^*D^*}M^4&=&{1\over
C_{D^*D^*}~f_{D^*D^*}(M^2)}
\Biggl[2f_\pi-{4m_c\langle\bar{q}q\rangle\over
3f_\pi M^2}
\nonumber\\
&-&{f_\pi\delta^2\over3M^2}\Biggl({23\over3}+{5m_c^2\over3M^2}\Biggr)\Biggr]
{e^{-m_c^2/M^2}\over M^2}\,,
\eeqa
\beqa
\Lambda_5^{D^*D^*}+A_5^{D^*D^*}M^2+B_5^{D^*D^*}M^4={1\over
C_{D^*D^*}~f_{D^*D^*}(M^2)}
\Biggl[-f_\pi+{23f_\pi\delta^2\over18M^2}\Biggr]
{e^{-m_c^2/M^2}\over M^2}\,,
\eeqa
\beqa
\Lambda_6^{D^*D^*}+A_6^{D^*D^*}M^2+B_6^{D^*D^*}M^4={1\over
C_{D^*D^*}~f_{D^*D^*}(M^2)}
\Biggl[-f_\pi+{25f_\pi\delta^2\over18M^2}\Biggr]
{e^{-m_c^2/M^2}\over M^2}\,,
\eeqa
\beqa
\Lambda_7^{D^*D^*}+A_7^{D^*D^*}M^2+B_7^{D^*D^*}M^4&=&{1\over
C_{D^*D^*}~f_{D^*D^*}(M^2)}
\Biggl[f_\pi-{2m_c\langle\bar{q}q\rangle\over
3f_\pi M^2}
\nonumber\\
&-&{3m_cf_{3\pi}\over M^2}\Biggr]
{e^{-m_c^2/M^2}\over M^2}\,,
\eeqa
\beqa
\Lambda_8^{D^*D^*}+A_8^{D^*D^*}M^2+B_8^{D^*D^*}M^4&=&{1\over
C_{D^*D^*}~f_{D^*D^*}(M^2)}
\Biggl[f_\pi-{2m_c\langle\bar{q}q\rangle\over
3f_\pi M^2}
\nonumber\\
&+&{2m_cf_{3\pi}\over M^2}\Biggr]
{e^{-m_c^2/M^2}\over M^2}\,,
\eeqa
\beqa
\Lambda_9^{D^*D^*}+A_9^{D^*D^*}M^2+B_9^{D^*D^*}M^4&=&{1\over
C_{D^*D^*}~f_{D^*D^*}(M^2)}
\Biggl[-{m_c^2f_\pi\over2}+{m_c\langle\bar{q}q\rangle\over
2f_\pi}
\nonumber\\
+{f_\pi\delta^2\over3}\Biggl({53\over6}&-&{13m_c^2\over6M^2}
+{5m_c^4\over12M^4}\Biggr)\Biggr]
{e^{-m_c^2/M^2}\over M^2}\,,
\eeqa
\beqa
\Lambda_{10}^{D^*D^*}+A_{10}^{D^*D^*}M^2+B_{10}^{D^*D^*}M^4={1\over
C_{D^*D^*}~f_{D^*D^*}(M^2)}
\Biggl[{2m_c\langle\bar{q}q\rangle\over3f_\pi}
+m_cf_{3\pi}\Biggr]
{e^{-m_c^2/M^2}\over M^4}\,,
\eeqa
\beqa
\Lambda_{11}^{D^*D^*}+A_{11}^{D^*D^*}M^2+B_{11}^{D^*D^*}M^4&=&{2\over
C_{D^*D^*}~f_{D^*D^*}(M^2)}
{m_c\langle\bar{q}q\rangle\over f_\pi}{e^{-m_c^2/M^2}\over M^2}\,,
\eeqa
\beqa
\Lambda_{12}^{D^*D^*}+A_{12}^{D^*D^*}M^2+B_{12}^{D^*D^*}M^4={1\over
C_{D^*D^*}~f_{D^*D^*}(M^2)}
\Biggl[2m_cf_{3\pi}-{f_\pi\delta^2\over3}\Biggr]
{e^{-m_c^2/M^2}\over M^4}\,,
\eeqa
\beqa
\Lambda_{13}^{D^*D^*}+A_{13}^{D^*D^*}M^2+B_{13}^{D^*D^*}M^4=-{2\over
C_{D^*D^*}~f_{D^*D^*}(M^2)}{f_\pi\delta^2\over3}
{e^{-m_c^2/M^2}\over M^4}\,,
\eeqa

where
\beq
f_{D^*D^*}(M^2)= {1\over\mds^2-\mpsi^2}\Biggl[{e^{-\mds^2/
M^2}\over M^2}- {e^{-\mds^2/M^2}-
e^{-\mpsi^2/M^2} \over\mpsi^2-\mds^2}\Biggr]
\eeq
and 
\beq
C_{D^*D^*}={1\over\mds^2\mpsi f_{D^*}^2f_\psi}\,.
\eeq

\section{Sum rules for the meson decay constants}

For consistency we use in our analysis the QCDSR expressions for 
the decay constants of the $J/\psi,~D^*$ and $D$ mesons
up to dimension four in lowest order of $\alpha_s$:
\beqa
&&f_D^2 = {3m_c^2\over 8\pi^2m_D^4}\int_{m_c^2}^{s_D}ds~ 
{(s-m_c^2)^2\over s}e^{(m_D^2-s)/M^2}\,
\nonumber\\
&&\,- {m_c^3\over m_D^4}
\langle\bar{q}q\rangle e^{(m_D^2-m_c^2)/M^2}\; ,\label{fhr} \\
&&f_{D^*}^2 = {1\over 8\pi^2\mds^2}\int_{m_c^2}^{s_{D^*}}ds~ 
\Biggl[{(s-m_c^2)^2
\over s}
\left(2+{m_c^2\over s}\right) \nonumber\\
&&\,\times\, e^{(\mds^2-s)/M^2}\Biggr]
\,-\,{ m_c\over \mds^2}\langle\bar{q}q\rangle e^{(\mds^2-m_c^2)/M^2},
\label{fhs} \\
&&f_\psi^2 = {1\over4\pi^2}\int_{4m_c^2}^{s_\psi}ds~{(s+2m_c^2)
\sqrt{s-4m_c^2}\over s^{3/2}}e^{(\mpsi^2-s)/M^2},\label{fpsi}
\eeqa
where $s_M$ stands for the continuum threshold of the meson $M$, which
we parametrize as $s_M=(m_{M}+\Delta_s)^2$. 
The values of $s_M$ are, in general, extracted from the two-point
function sum rules for $f_D$ and $f_{D^*}$ and $f_\psi$ in Eqs.~(\ref{fhr}), 
(\ref{fhs}) and (\ref{fpsi}). Using the Borel region 
$3 \leq M_M^2\leq 6 \GeV^2$ for the $D^*$ and  $D$ mesons and 
$6 \leq M_M^2 \leq 12 \GeV^2$ for the $J/\psi$, we found good stability 
for $f_D$,  $f_{D*}$ and $f_\psi$ with $\Delta_s\sim0.6\GeV$. We obtained 
$f_D=160\pm5\MeV$, $f_{D^*}=220\pm10\MeV$ and $f_\psi=280\pm10\MeV$, which
are compatible with the numerical values in Eq.~(\ref{num}).


\begin{references}

\bibitem{ma86} T. Matsui and H. Satz, Phys. Lett. {\bf B178}, 416 (1986).

\bibitem{thews} R. L. Thews,  hep-ph/0206179; R. L. Thews and J.   
Rafelski, Nucl. Phys. {\bf A698}, 575 (2002); 
R. L. Thews, M. Schroedter, J. Rafelski  Phys. Rev. {\bf C63}, 054905 (2001). 

\bibitem{rapp} L. Grandchamp and R. Rapp, hep-ph/0209141;  
      Nucl. Phys. {\bf A709}, 415 (2002);  Phys. Lett. {\bf B523},  60  (2001).

\bibitem{huf} B.Z. Kopeliovich, A. Polleri and J. H\"ufner,  Phys. Rev. Lett. 
{\bf 87},  112302 (2001), and references therein.

\bibitem{dnn}  F.O. Dur\~aes, F.S. Navarra and M. Nielsen,   nucl-th/0210043.

\bibitem{bhp} G. Bhanot and M.E. Peskin, Nucl. Phys. {\bf B156}, 391  
              (1979); M.E. Peskin, Nucl. Phys. {\bf B156}, 365  (1979).

\bibitem{kha2} D. Kharzeev and H. Satz, Phys. Lett. {\bf B334}, 155  (1994).

\bibitem{lo} S.H.Lee and Y. Oh, J.Phys. {\bf G28}, 1903 (2002);  
             Y. Oh, S. Kim, S.H. Lee, Phys. Rev.  {\bf C65}, 067901 (2002). 

\bibitem{wongs} C.-Y. Wong, E.S. Swanson and T. Barnes, Phys. Rev.
                {\bf C62}, 045201 (2000); {\bf C65}, 014903 (2001).

\bibitem{mbq} K. Martins, D. Blaschke and E. Quack, Phys. Rev.
              {\bf C51}, 2723 (1995).

\bibitem{mamu98} S.G. Matinyan and B. M\"uller, Phys. Rev. {\bf C58}, 
                 2994 (1998).

\bibitem{osl} Y. Oh, T. Song and S.H. Lee, Phys. Rev. {\bf C63}, 
              034901 (2001). 


\bibitem{haglin} K.L. Haglin, Phys. Rev. {\bf C61}, 031902 (2000);
                 K.L. Haglin and C. Gale, Phys. Rev. {\bf C63}, 065201 (2001).


\bibitem{linko} Z. Lin and C.M. Ko,  Phys. Rev. {\bf C62}, 034903 (2000). 


\bibitem{ikbb} V.V. Ivanov, Yu.L. Kalinovsky, D.B. Blaschke and G.R.G. Burau,
               hep-ph/0112354.


\bibitem{nnr} F.S. Navarra, M. Nielsen and M.R. Robilotta, Phys. Rev. 
              {\bf C64}, 021901(R) (2001).


\bibitem{kardok} D. Kharzeev, Nucl. Phys. {\bf A638}, 279c (1998).


\bibitem{hdnnr} H.G. Dosch,  F.S. Navarra,  M. Nielsen and M. Rueter, 
                Phys. Lett. {\bf B466},  363 (1999). 


\bibitem{muel} B. Mueller, Nucl. Phys. {\bf A661}, 272 (1999).

\bibitem{zinovjev} K. Redlich, H. Satz, G. M. Zinovjev, Eur. Phys. J. 
{\bf C17},  461 (2000).

\bibitem{hufkop} J. Huefner, B.Z. Kopeliovich, Phys. Lett. {\bf B 426},  154 
(1998).

\bibitem{nnmk} F.S. Navarra, M. Nielsen, R.S. Marques de Carvalho and G. Krein,
               Phys. Lett. {\bf B529}, 87 (2002).

\bibitem{oslw}  Y. Oh, T. Song, S.H. Lee and C.-Y. Wong, nucl-th/0205065.

\bibitem{gerland} L. Gerland, L. Frankfurt, M. Strikman, H. Stoecker and W.
Greiner,  Phys. Rev. Lett. {\bf 81}, 762 (1998).

\bibitem{nosform} F.S. Navarra and M. Nielsen, 
                  Phys. Lett. {\bf B443},) 285 (1998);
                  F.O. Dur\~aes, F.S. Navarra and M. Nielsen, 
                  Phys. Lett. {\bf B 498}, 169 (2001);  
      M.E. Bracco, M. Chiapparini, A. Lozea, F.S. Navarra and M. Nielsen,  
                  Phys. Lett. {\bf B 521}, 1 (2001); 
                  F.S. Navarra, M. Nielsen and M.E. Bracco, 
                  Phys. Rev. {\bf D 65}, 037502 (2002);  
     R.D. Matheus, F.S. Navarra, M. Nielsen and R. Rodrigues da Silva, 
                  Phys. Lett. {\bf B541}, 265  (2002). 

\bibitem{nos} F.S. Navarra, M. Nielsen, M.E. Bracco, M. Chiapparini and
              C.L. Schat, Phys. Lett.  {\bf B489},  319  (2000). 

\bibitem{bnn} M.E. Bracco, F.S. Navarra and M. Nielsen, 
              Phys. Lett. {\bf B454}, 346 (1999).
%
\bibitem{svz}  M.A. Shifman, A.I. Vainshtein and V.I. Zakharov, Nucl. 
Phys.  {\bf B120}, 316 (1977).
%
\bibitem{rry} L.J. Reinders, H. Rubinstein and S. Yazaki, Phys.
Rep. {\bf 127}, 1 (1985). 
%
\bibitem{kl}  H. Kim and S.H. Lee, Eur. Phys. Jour. {\bf C22}, 707 (2002).
%
\bibitem{klo}  H. Kim, S.H. Lee and M. Oka, Phys. Lett.  {\bf B453},  199 
(1999). 
%
\bibitem{bbk}
V.M.~Belyaev, V.M.~Braun, A.~Khodjamirian and R.~Ruckl,
Phys.\ Rev.\ D {\bf 51}, 6177 (1995).
%
%
\bibitem{io2}  B.L. Ioffe and A.V. Smilga, Nucl. Phys. {\bf B232}, 109
(1984).
%
\bibitem{dlnn} F.O. Dur\~aes, S.H. Lee, F.S. Navarra, M. Nielsen, 
nucl-th/0210075.

\bibitem{arleo} F. Arleo, P.-B. Gossiaux, T. Gousset and J. Aichelin, 
                Phys. Rev. {\bf D65}, 014005 (2002).

\end{references}
\end{document}